\documentclass[%
 reprint,
 superscriptaddress,
 amsmath,amssymb,
 aip,
 apl,
floatfix,
]{revtex4-1}

\usepackage{graphicx}
\usepackage{color,soul}
\usepackage[colorlinks=true,linkcolor=blue]{hyperref}
\usepackage{dcolumn}
\usepackage{bm}

\begin{document}

\title{Steering light with magnetic textures}

\author{Ioan-Augustin Chioar}
\altaffiliation{Current address: Department of Applied Physics, Yale University, New Haven 06511, CT, USA.}
\affiliation{Department of Physics and Astronomy, Uppsala University, Box 516, SE-75120 Uppsala, Sweden}

\author{Christina Vantaraki}
\affiliation{Department of Physics and Astronomy, Uppsala University, Box 516, SE-75120 Uppsala, Sweden}

\author{Merlin Pohlit}
\affiliation{Department of Physics and Astronomy, Uppsala University, Box 516, SE-75120 Uppsala, Sweden}

\author{Richard M.  Rowan-Robinson}
\altaffiliation{Current address: Department of Material Science and Engineering, University of Sheffield, Sheffield, United Kingdom.}
\affiliation{Department of Physics and Astronomy, Uppsala University, Box 516, SE-75120 Uppsala, Sweden}

\author{Evangelos Th. Papaioannou}
\affiliation{Institut f\"ur Physik, Martin-Luther-Universit\"at Halle-Wittenberg,  Von-Danckelmann-Platz 2, 06120 Halle, Germany}

\author{Bj\"orgvin Hj\"orvarsson}
\affiliation{Department of Physics and Astronomy, Uppsala University, Box 516, SE-75120 Uppsala, Sweden}

\author{Vassilios Kapaklis}
\email[Corresponding author: ]{vassilios.kapaklis@physics.uu.se}
\affiliation{Department of Physics and Astronomy, Uppsala University, Box 516, SE-75120 Uppsala, Sweden}

\begin{abstract}
We study the steering of visible light using a combination of magneto-optical effects and the reconfigurability of magnetic domains in Yttrium-Iron Garnet films. The spontaneously formed stripe domains are used as a field-controlled optical grating, allowing for active spatiotemporal control of light. We discuss the basic ideas behind the approach and provide a quantitative description of the field dependence of the obtained light patterns. Finally, we calculate and experimentally verify the efficiency of our magneto-optical grating.
\end{abstract}

\maketitle

Optical components are used for focusing, filtering, steering and manipulating the polarization of light. Their properties are typically obtained by the fabrication of three dimensional objects, having specific refractive indices, dichroism or birefringence effects\cite{Chen_NatRevMat_2020}. However, the mechanical assembly of re-configurable components forming an optical device can limit its long-term stability and reliability. Flat optics has therefore been pursued, in an attempt to remedy these shortcomings. The effort has given rise to a revolution in the field of optics, building upon developments in the fields of plasmonics, metamaterials and nanofabrication \cite{Optical_computing_NatPhot2019,Shaltout_spatiotemporal,Shaltout:2019bc, rho_2020}.  Metamaterials facilitate the shaping of optical wavefronts  \cite{Yu_Capasso_Science_2011,Yu_Capasso_NatMat_2014,Chen_NatRevMat_2020} through the structuring of near-fields in a designer manner, thereby offering control of the far-field response \cite{Ginis_Science_2020}. These can also be reconfigured, using electric and magnetic fields, temperature, mechanical as well as chemical reactions\cite{shalaginov_design_2020}. Magnetically controlled metamaterials are of special relevance in this context, due to their reconfigurability, flexibility in design and the fast response of opto-magnetic effects \cite{Nanoscale_Magnetophotonics_JAP2020, zvezdin2020modern}.
Therefore, investigating ways in which tailored magnetic textures can be harnessed for the design of useful optical responses is of particular interest \cite{zvezdin2020modern, CostaKramer_2003,syouji_magneto-optical_2013,mito_stress_2018,higashida_diffraction_2020}.In fact, related approaches for magnetic holography recording and the subsequent steering of light were initiated in the 1970s using garnet materials \cite{Johansen_1971,lacklison_garnets_1973,Scott_1976,Sauter_1977,krawczak_torok_1980,numata_stripe_1980,sauter_alterable_1981,hansen_magnetic_1984}, MnBi alloys\cite{MnBi_1968,Mezrich_1969,rull_recording_1976} and EuO \cite{Fan_Hologram_1969}.

In light of the metamaterial approach and opportunities, we here revisit the use of Yttrium-Iron Garnet (YIG) \cite{Johansen_1971, Scott_1976, sauter_alterable_1981} for obtaining one of the most basic functions of an optical component: deflection of light. The steering is obtained by applying an external magnetic field, influencing the spontaneously formed stripe-like magnetic domains, which in turn affect the intensity of the angular distribution of the transmitted light. We describe the connection between the magnetic texture of the YIG film and the deflected light, establishing a description of the relation between the real space magnetic domain structure and the reciprocal space, as seen from the scattered light patterns\cite{schmitte_bragg-moke_2003,Vavassori_BraggMOKE_2004, Vavassori_PRB_2004, Arnalds_BraggMOKE_2010}. We also quantify the efficiency of the YIG-based grating. The ideas and results discussed here can be utilized for the design and development of a new generation of flat, reconfigurable and potentially fast optics\cite{Nanoscale_Magnetophotonics_JAP2020} through the control of magnetic order and textures at the mesoscale\cite{CostaKramer_2003} using thin film technology and nanolithography \cite{Wang_Nature_2006, Perrin_Nature_2016,Nisoli_2017_NatPhys,Ostman_NatPhys_2018,Rougemaille_review_2019, skjaervo_advances_2020}.

\begin{figure}[ht!]
    \centering
    \includegraphics[width=\columnwidth]{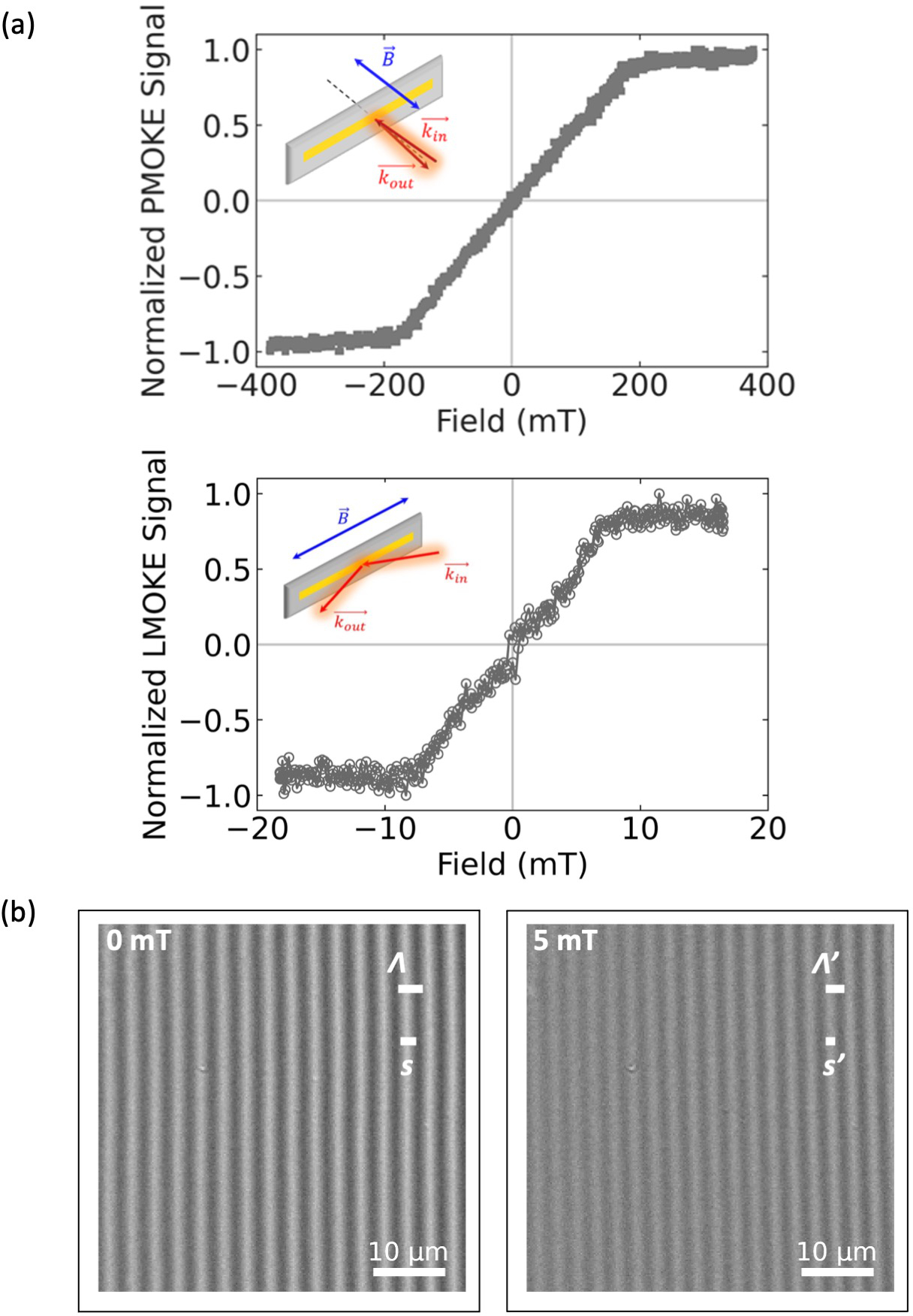}
    \caption{The magneto-optical grating. (a) Hysteresis loops of the YIG sample for out-of-plane (upper panel) and in-plane (lower panel) applied magnetic fields.
    (b) An in-plane magnetic field couples to the in-plane component of the magnetization in the YIG film, altering the periodicity $\Lambda$ while maintaining the ratio $f=s/\Lambda=s'/\Lambda'$ between the domains. 
    The full micromagnetic characterization of the YIG film, using magnetic microscopy is detailed in the Supplementary Videos.}
    \label{fig:Figure1}
\end{figure}

The investigated YIG film ($30\,\rm{mm}\times 3\,\rm{mm}\times 7.3\,\rm{\mu m}$) was grown by liquid phase epitaxy on a 500 $\mu$m thick Gallium Gadolinium Garnet (GGG) substrate \cite{YIG_KL_APL_2014,Schmidt:2020de}. The angular dependence of the transmitted intensity was determined using a specially designed magneto-optical diffractometer, based on a $\theta-2\theta$ goniometer (Huber 424 2-circle goniometer). The sample was mounted in the center of a quadrupole magnet, providing vectorial magnetic fields up to 42~mT. The sample was illuminated using a supercontinuum laser (Fianium SC-400-2), with a wavelength range of 400 - 1100~nm or a monochromatic laser (Coherent OBIS) with a wavelength of 530~nm and power of 20~mW. Two Glan-Thompson polarizers (Thorlabs GTH10M) were used for setting the polarization of the incoming beam and for analyzing the rotation in the detected light. The signal was modulated to allow lock-in detection (SR830) and recorded using a Si photodiode detector (Thorlabs DET100A). A beam-splitter was employed for monitoring the intensity of the source, providing on-the-fly normalization of the intensity of the incoming light. For the field dependence measurements, the sample was first saturated, ensuring a reset of the magnetic domain configuration, and thereafter brought to the targeted field before performing a detector scan. Hysteresis curves were measured for both in- and out-of-plane applied magnetic fields using a magneto-optical Kerr effect (MOKE) magnetometer. Finally, a Kerr microscope was used for magnetic imaging. To image the remanent magnetization state, the samples were first demagnetized in a time-dependent magnetic field of decaying amplitude. The microscope data presented here are polar-MOKE (P-MOKE) contrast images in reflection.

\begin{figure}[ht!]
    \centering
    \includegraphics[width=\columnwidth]{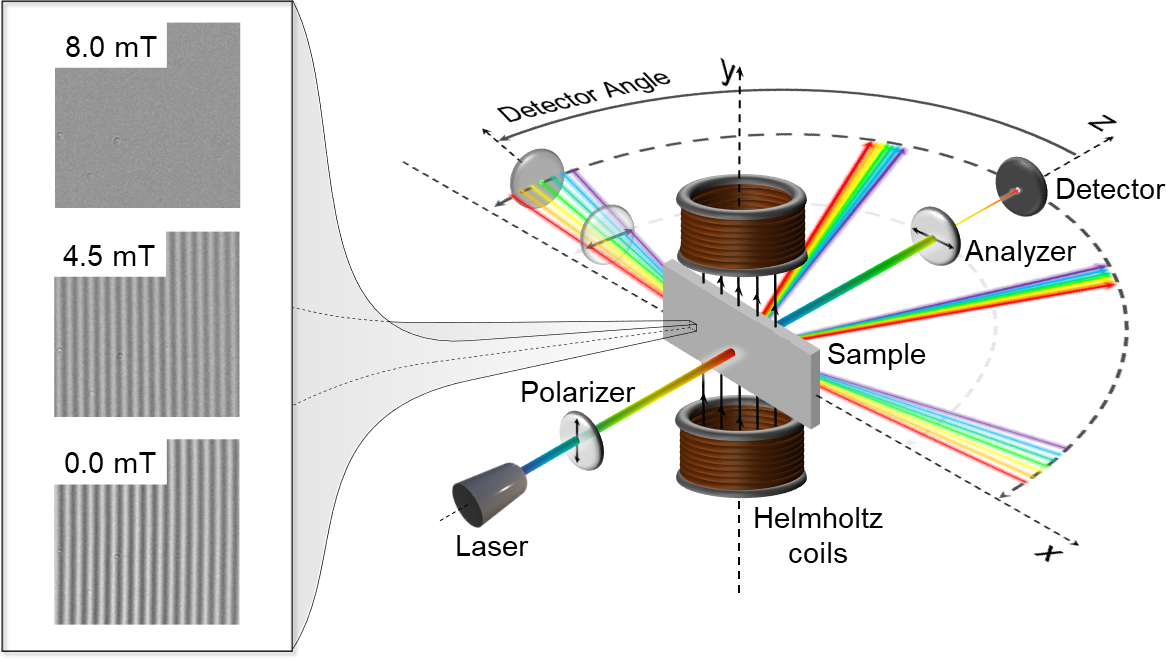}
    \caption{Schematic description of the magneto-optical scattering setup and the field dependence of the  binary magnetic YIG grating. The YIG film is illuminated using a supercontinuum laser beam at normal incidence (450 - 900 nm), while placed between two Helmholtz coils, providing a field along the $y$-direction. The laser beam is linearly polarized before reaching the sample. An analyzer is positioned in front of the detector, which can be moved in the $zx$ plane, as depicted on the right side. }
    \label{fig:Figure2}
\end{figure}

Magnetic stripe domains are formed in the YIG film as shown in Fig. \ref{fig:Figure1}, constituting a one-dimensional grating-like structure for the out-of-plane magnetization component. The stripe domains can be oriented\cite{Rimantas} along any direction within the sample plane, using external in-plane magnetic fields (see also Supplementary Videos). The direction of the applied in-plane magnetic field also affects the magnetic texture by primarily altering the grating periodicity ($\Lambda$) (Fig. \ref{fig:Figure1}b), with the primary domain width (s) selectively affected, depending on the in-plane field direction and magnitude\cite{zvezdin2020modern}. For the remainder of this letter, we will concentrate on the case where in-plane magnetic fields are applied to the YIG film along the $y$-direction, as defined in Fig.~\ref{fig:Figure2}.

The layout of the magneto-optical scattering experimental setup is illustrated in Fig.~\ref{fig:Figure2}.
The Faraday effect acting upon the light (polarized along the $y$-direction) transmitted through the YIG film results in a rotation of the polarization. Having domains of opposite out-of-plane magnetization components yields rotations of opposite signs. The interference between the light with opposite rotation of the polarization gives rise to a diffraction pattern, closely resembling that of a conventional optical grating. However, the interference arises from the phase difference of the partial waves and not a modulation in transmitted intensity along the grating direction. Assuming a linear-response regime, we can further calculate the intensity of the transmitted diffracted beam though the YIG film. Defining $F$ as the Faraday rotation and $d$ as the film thickness, the rotation will be $\varphi = Fd$. Domains of opposite magnetization, rotate the polarization in opposite directions ($\varphi_1=+Fd$ for $M^+$ and $\varphi_2=-Fd$ for $M^-$), resulting in a periodic modulation of the electric field components. Consequently, a maximum achievable efficiency in terms of change in beam power can be estimated, knowing the attenuation coefficient $a$ and by using (see Supplemental Material for full derivation)\cite{Haskal_1970}:

\begin{equation}
    \eta_{max} = \frac{4}{\pi^2} e^{-2} \left( \frac{2F}{a} \right)^2 sin^2\frac{\pi s}{\Lambda}
\end{equation}

For the YIG film used here: $F = 2200$ deg/cm (experimentally determined, see Supplemental Material), $a$~=~1417~cm$^{-1}$ (measured absorption coefficient, see Supplemental Material), resulting in $\eta_{max} = 1.6\times10^{-4}$, for a wavelength of $\lambda$ = 530 nm. This value is comparable to, yet higher than certain reported results in the literature, for example MnBi magnetic gratings\cite{Mezrich_1969,MnBi_1968,tanaka_diffraction_1972,rull_recording_1976}, yet smaller than the values reported for Bi-substituted garnet materials\cite{lacklison_garnets_1973,Scott_1976,Sauter_1977}. It is worth noting that these improvements can imply additional chemical synthesis complexity and an intricate interplay between the magnetic properties and film thickness which impacts the angular deflection window and magneto-optical efficiency\cite{lacklison_garnets_1973,Scott_1976}. The actual experimental value of the efficiency for our YIG film was determined to be $\eta_{exp}  = 1.47(6)\times10^{-4}$, in reasonable agreement with the calculated value. 

\begin{figure}[t!]
    \centering
    \includegraphics[width=\columnwidth]{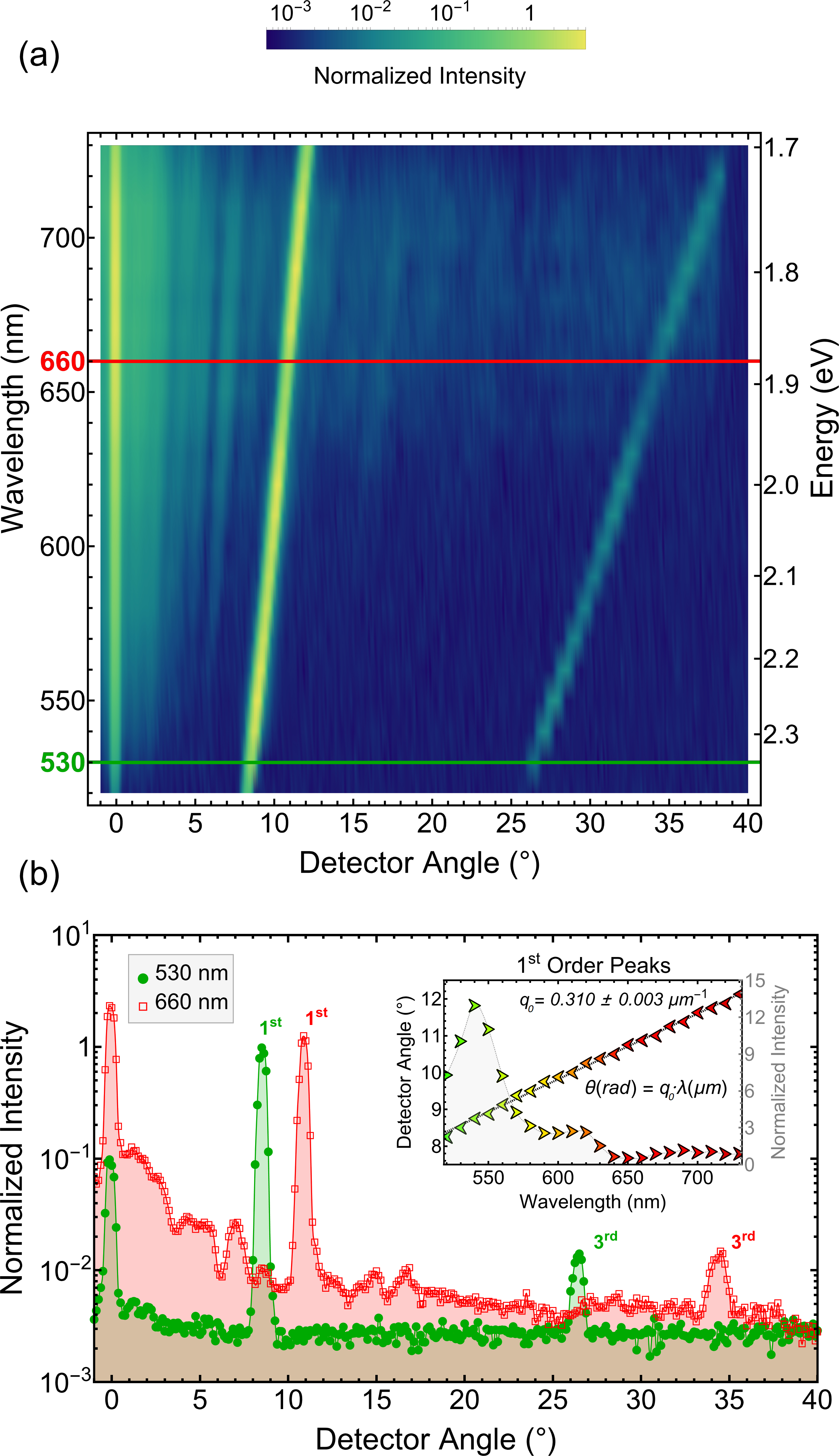}
    \caption{Wavelength dependence of the magneto-optical scattering and the remanent magnetic state, where $\Lambda \approx 3.2$~$\mu$m. (a) The diffraction peaks move when the laser wavelength is changed, as in conventional optical grating. (b) Fitting the angular position of the first order peak against the laser wavelength yields the reciprocal lattice unit value $q_0=q/(2\pi)=1/\Lambda$ (see Supplementary Material for details) shown in the inset, which closely follows the observed periodicity as observed by magneto-optical microscopy (Fig. \ref{fig:Figure1}). 
    }
    \label{fig:Figure3}
\end{figure}

\begin{figure}[t!]
    \centering
    \includegraphics[width=\columnwidth]{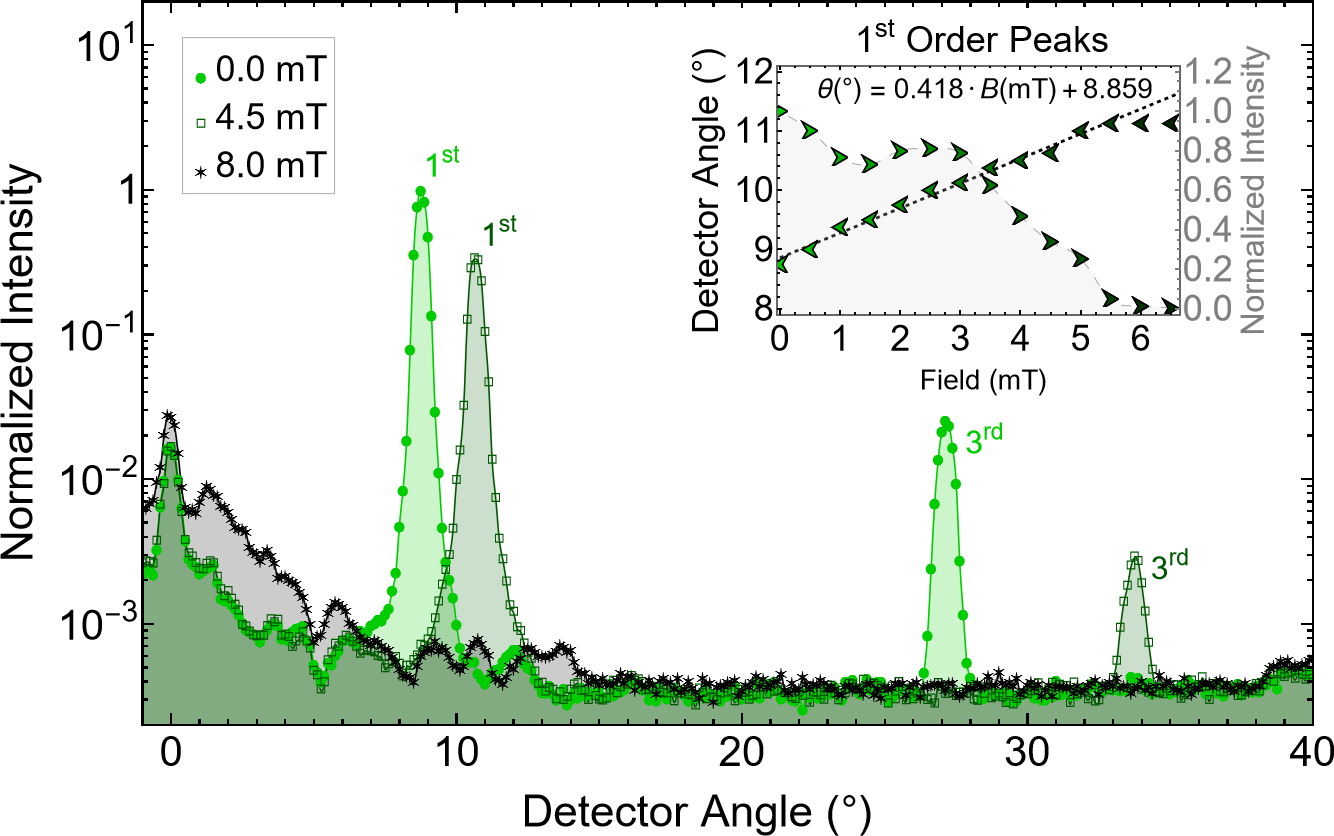}
    \caption{Field dependence of the magneto-optical scattering for in-plane applied fields along the $y$-direction (Fig. \ref{fig:Figure2}).
    All patterns were recorded using a wavelength of 530 nm. The inset shows the field dependence of the position and the intensity of the first diffraction peak. The intensities have been normalized to the first diffraction order in absence of an applied magnetic field (0 mT).}
    \label{fig:Figure4}
\end{figure}

Fig.~\ref{fig:Figure3}a displays the wavelength and angular dependence of the transmitted light 
and in Fig.~\ref{fig:Figure3}b we show the angular dependence of the intensity at two wavelengths (660 and 530 nm) along with the position and the intensity of the first diffraction peak. We note the close to perfect scaling of the angular position of the peak and the wavelength of the incoming light as well as the strong wavelength dependence of the intensity of the diffracted light, reminiscent of the YIG intrinsic magneto-optical activity (see Supplemental Material). The in-plane field dependence of the diffracted light is illustrated in Fig.~\ref{fig:Figure4}. As the applied field is increased, the grating periodicity decreases, leading to an increase of the diffraction angle for any given order, while a decrease of the intensity is also recorded. The latter can be traced to Fig. \ref{fig:Figure1}b, originating from a reduction in the P-MOKE contrast as the field increases.

The stripe domains disappear, as does the diffraction, when the sample is saturated. Starting from remanence and with the field applied parallel to the sample surface, a linear dependence of the angular position of the first order diffracted beam with the applied field strength is observed, almost the whole way up to magnetic saturation, as shown in the inset of Fig.~\ref{fig:Figure4}. At the same time the intensity of the diffracted beam is generally decreasing with the increase in applied field, ultimately reaching zero at magnetic saturation. The decrease in the diffracted intensity originates from a reduction in the out-of-plane magnetization, thus decreasing the difference in the rotation of the polarization angle in the stripe domains. Note that the azimuthal rotation of the in-plane applied field starting from saturation results in the rotation of the scattering plane, since the stripe domains form parallel to the new field direction (see Supplementary Video).


\begin{figure}[t!]
    \centering
    \includegraphics[width=\columnwidth]{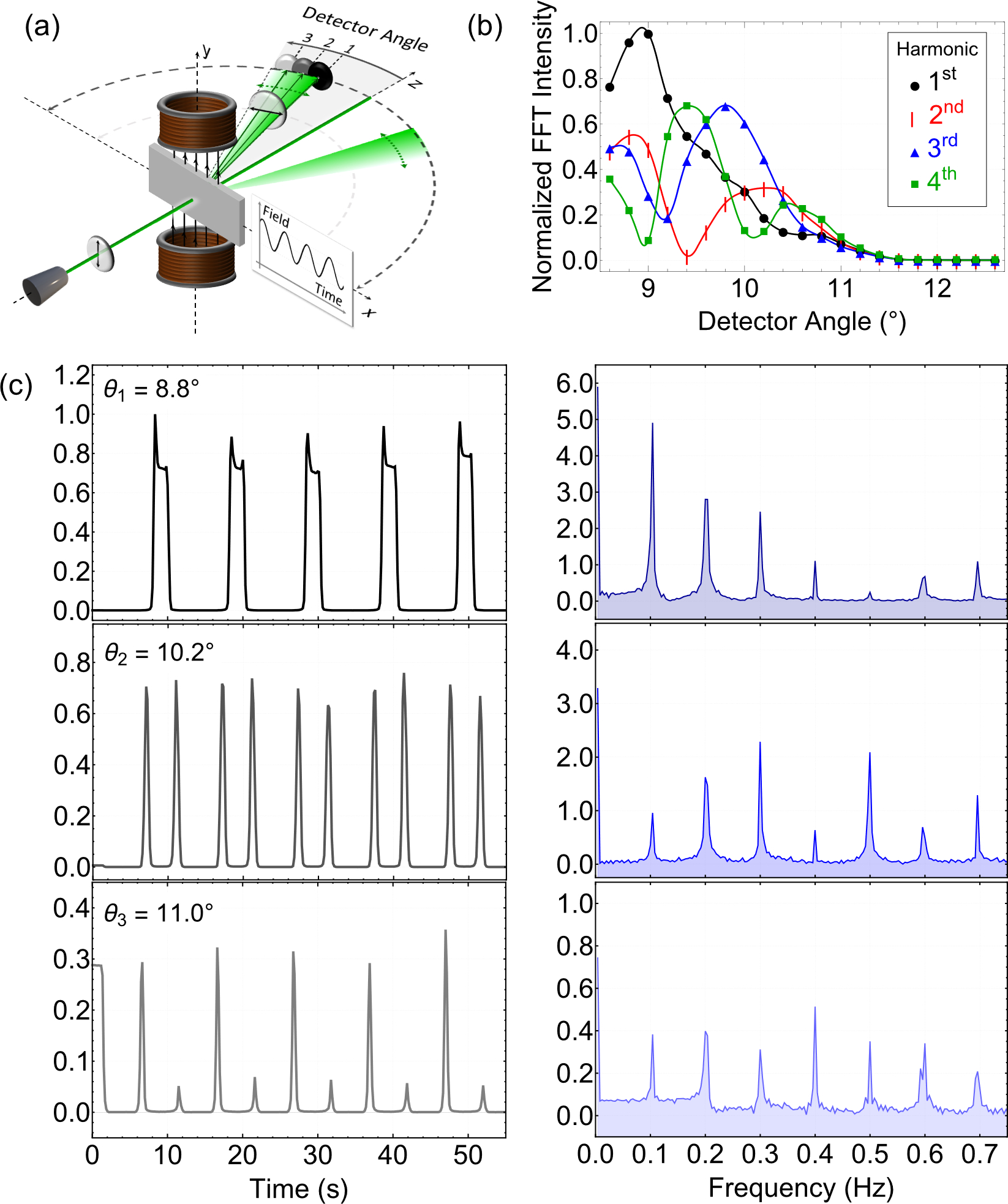}
    \caption{Frequency modulation using magneto-optical gratings. (a) Combining the detector angular placement with a field oscillation at a given amplitude and frequency, it is possible to alter the spectral intensity content of the signal as shown in (b) by changing the beam sweeping sequence over the detector (c). In all cases shown here, an oscillating field of 5~mT amplitude, 5~mT offset and 0.1~Hz frequency was used, with the laser wavelength set to 530~nm. The recorded intensities are for the  first diffraction order and detector angles of $\theta_1=8.8^{\circ}$, $\theta_2=10.2^{\circ}$ and $\theta_3=11.0^{\circ}$. The intensities have been normalized to the highest intensity of the first harmonic (0.1~Hz). This highlights the tunable harmonic content, enabling modulation on frequencies above the field driving frequency.} 
    \label{fig:Figure6}
\end{figure}

Finally, we describe the temporal response of the YIG magneto-optical diffraction. For this purpose, we used the experimental protocol illustrated in Fig.~\ref{fig:Figure6}a. The time-dependent response depends on the field protocol as well as the position of the detector. 
Having chosen a detector angle within the angular position window of the first order peaks, we applied a sine-wave magnetic field on top of a field offset, effectively driving the sample between its saturated and remanent states. This results in a time dependency, as exemplified in the left column of Fig.~\ref{fig:Figure6}c. Here we notice the large difference in response, solely arising from the choice of the detector angle. In fact, the relative positioning of the detector within the first order peak window produces signals with mixed spectral load and possibilities for beam modulation options involving tunable weighing of the harmonic content. Changing the amplitude and the sign of the applied field, i.e. performing partial or extended loops, is expected to add another degree of freedom, allowing further tailoring of the spatiotemporal steering of light. An interesting outlook also resides in the use of driving frequencies beyond the current quasi-static framework.


The concepts discussed here can be used when designing magnetically controlled flat optical devices. A foreseeable major potential for improvement lies in the field of magnetic metamaterials\cite{skjaervo_advances_2020, Nisoli_2017_NatPhys,Nanoscale_Magnetophotonics_JAP2020}, where the necessary magnetic domain structures can be designed and engineered utilizing lithography. For reasonably large diffraction angles to be achieved, the width of the domains must be comparable or larger than the wavelength of the light, for which nanopatterned magnetic metamaterials offer an ideal setting\cite{CostaKramer_2003}. This can be done in combination with conventional magnetic materials rather than targeting specific magnetic materials with the required, intrinsic domain structures, such as the YIG presented here. Additionally, a variety of magnetic materials suitable for the fabrication of such metamaterials exhibit all-optical switching properties, where ultra-fast laser pulses may be used to set the magnetic state in nanoarrays \cite{Rich_Truncated_nanocones, 2021_TbCo_Rings} or films of these materials\cite{Mangin_NatMat_2014, Ciuciulkaite_TbCo_AOS_2020, Ksenzov_2021}. In this way, light can be acted upon by metamaterial architectures, but also be used to set this action by {\it writing-in} the necessary mesoscopic magnetic structure. Advanced design and control of such metamaterials will allow for more intricate schemes of light control\cite{Hasman_NNano_2020, Ksenzov_2021} in terms of the scattering, but also over properties of light such as angular and orbital momenta\cite{Beth:1935jg,Beth:1936cm,Allen:1992by,Woods_OAM_2020arXiv201110148W}, holding strong promises for information technology related applications. 

\section*{Supplementary Material}
Supplemental material includes a detailed derivation of the magneto-optical grating equations and efficiency, along with experimental data on the photon energy-dependent Faraday rotation and absorption coefficient. We further present details on the calculation of the scattering patterns for real-space magnetic microscopy data. Two videos are included, presenting the field dependence of the magnetic domain structure alongside the resulting reciprocal space patterns, and the experimentally observed light beam deflections while applying magnetic fields.

\begin{acknowledgements}
The authors acknowledge support from the Knut and Alice Wallenberg Foundation (project 2015.0060), STINT (project, KO2016-6889) and the Swedish Research Council (project 2019-03581). The authors would like to express their gratitude to Prof. G. Andersson for providing guidance with the Kerr microscopy measurements. V.K. would like to thank Prof. P. M. Oppeneer and Prof.~Alexandre Dmitriev for fruitful discussions.
\end{acknowledgements}

\section*{Data availability}
The data that support the findings are available from the corresponding authors upon request.

\section*{References}

%

\pagebreak
\onecolumngrid
\newpage
\begin{center}
\textbf{\large Supplemental Material: Spatiotemporal steering of light using magnetic textures}
\end{center}
\setcounter{equation}{0}
\setcounter{figure}{0}
\setcounter{table}{0}
\setcounter{page}{1}
\makeatletter
\renewcommand{\theequation}{S\arabic{equation}}
\renewcommand{\figurename}{Supplementary FIG.}
\renewcommand{\thefigure}{{\bf \arabic{figure}}}
\renewcommand{\bibnumfmt}[1]{[S#1]}
\renewcommand{\citenumfont}[1]{S#1}
\renewcommand{\thepage}{S-\arabic{page}}

\section*{Equations of the magneto-optical grating}

\begin{figure}[b!]
    \centering
    \includegraphics[width=0.75\columnwidth]{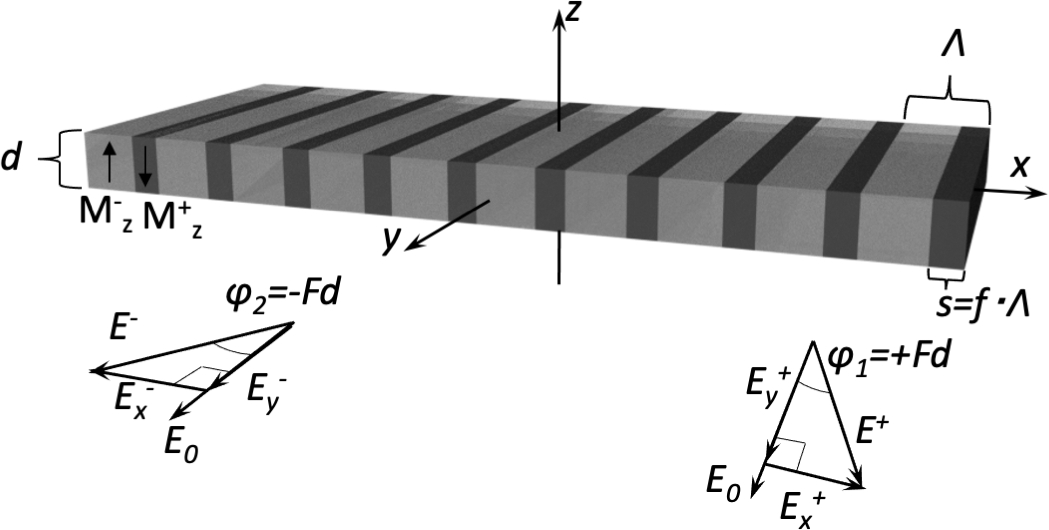}
    \caption{Schematic representation of a binary magnetic grating with a periodicity of $\Lambda$ and fraction $f=s/\Lambda$.}
    \label{fig:grating}
\end{figure}

We construct the magnetic domain grid by using a repetition of the unit box function $U(x)$, defined as

\begin{equation}
    U(x) = \begin{cases}1, & x \in [-\frac{1}{2},\frac{1}{2}]  \\ 0, & x \notin [-\frac{1}{2},\frac{1}{2}]\end{cases}
    \vspace{2mm}
\end{equation}
which we can further generalize to include an arbitrary offset ($x_0$) and an arbitrary box width ($\Lambda$):

\begin{equation}
    U\left(\frac{x-x_0}{\Lambda}-\frac{1}{2}\right) = \begin{cases}1, & x \in [x_0,x_0+\Lambda]  \\ 0, & x \notin [x_0,x_0+\Lambda]\end{cases}, \Lambda > 0
    \vspace{2mm}
\end{equation}

We can now write the grid as an offset sum of the base, which we in turn define as an asymmetric square wave, with widths of $f \Lambda$ and $(1-f) \Lambda$, respectively:

\begin{equation}
    G(x,\Lambda,f,N) = \sum_{m=0}^{N-1}U\left(\frac{x-m\Lambda}{f\Lambda}-\frac{1}{2}\right) - \sum_{m=0}^{N-1}U\left(\frac{x-(m+f)\Lambda}{(1-f)\Lambda}-\frac{1}{2}\right)
    \label{grating}
    \vspace{2mm}
\end{equation}
The fraction $f = \frac{s}{\Lambda} \in (0,1)$, in accordance with Supplementary Fig. \ref{fig:grating}. Fourier transforming Equation \ref{grating} with respect to x yields:

\begin{equation} 
\begin{split}
\mathcal{G}(q,\Lambda,f,N) & = \sum_{m=0}^{N-1}\left(\frac{2}{q} \cdot e^{iq\Lambda f \left( \frac{m}{f}+\frac{1}{2}\right)} \cdot sin\left( \frac{q\Lambda f}{2} \right) \right) \\
 & =  \sum_{m=0}^{N-1}\left(\frac{2}{q} \cdot e^{iq\Lambda (1-f) \left( \frac{m+f}{1-f}+\frac{1}{2}\right)} \cdot sin\left( \frac{q\Lambda (1-f)}{2} \right) \right)
\end{split}
\vspace{3mm}
\end{equation}
To further simplify, we introduce $q_0 = \frac{q}{2\pi}$ and therefore arrive at: 

\begin{equation}
    \mathcal{G}(q_0,\Lambda,f,N) = 
    \frac{1}{\pi q_0} \cdot e^{i \pi q_0 \Lambda f} \cdot \left(\frac{e^{i 2 \pi q_0 \Lambda N}-1}{e^{i 2 \pi q_0 \Lambda}-1}\right) \cdot \left[
    sin(\pi q_0 \Lambda f)-e^{i \pi q_0 \Lambda} \cdot sin(\pi q_0 \Lambda(1-f)) \right]
    \vspace{2mm}
\end{equation}
The light intensity $I$ will be proportional to:

\begin{equation}
        \mathcal{I}(q_0, \Lambda, f, N) = \mathcal{G} \cdot \mathcal{G}^* = \frac{1}{\pi^2 q_0^2} \cdot \Bigg(\frac{sin^2(\pi q_0 \Lambda N)}{sin^2(\pi q_0 \Lambda)} \Bigg) \cdot \\
        \Big[ sin^2(\pi q_0 \Lambda f) + 
        sin^2\big[\pi q_0 \Lambda (1-f)\big] - 2 \cdot sin(\pi q_0 \Lambda f) \cdot sin\big[\pi q_0 \Lambda (1-f)\big] \cdot cos(\pi q_0 \Lambda) \Big]
    \label{Intensity}
    \vspace{2mm}
\end{equation}
\vspace{2mm}

From Equation \ref{Intensity}, the zeroth order ($q_0=0$) diffraction peak intensity is given by:

\begin{equation}
    \mathcal{I}_0 (q_0=0, \Lambda, f, N) =  \Lambda^2 N^2 \cdot (1-2f)^2 = \mathcal{I}_{in} \cdot (1-2f)^2
    \vspace{2mm}
\end{equation}
where $\mathcal{I}_{in} = \Lambda^2 N^2$ can be considered as corresponding to an initial/incident intensity. As expected, for the case of a symmetric base square wave ($f = \frac{1}{2}$), the zeroth order diffracted intensity vanishes, $\mathcal{I}_0 (0, \Lambda, f = \frac{1}{2}, N)=0$, along with all even orders.
\vspace{5mm}

In a similar fashion we can estimate the intensity of the first order beam ($q_0=\frac{1}{\Lambda}$), which is proportional to:

\begin{equation}
    \mathcal{I}_1 (q_0=\frac{1}{\Lambda}, \Lambda, f, N) = \left[ \frac{2 \Lambda N}{\pi} \cdot sin(\pi f)\right]^2 = \mathcal{I}_{in} \cdot \bigg(\frac{2}{\pi}\bigg)^2 sin^2(\pi f)
    \vspace{2mm}
\end{equation}

\vspace{5mm}
    
Taking now into account the polarization profiles for the $x$- and $y$-directions and assuming a normal incidence $y$-polarized beam onto a sample of thickness $d$, Faraday rotation coefficient $F$ and absorption coefficient $\alpha$, we get:

\begin{equation*}
    I_0 = I_0^{x} + I_0^{y},
\end{equation*}

with

\begin{equation} 
\begin{split}
        I_0^{y} & = I_{in} \cdot e^{-\alpha d} \cdot cos^2(Fd), \\
        I_0^{x} & = I_{in} \cdot (1-2f)^2 \cdot e^{-\alpha d} \cdot sin^2(Fd) \\
        I_1 & = I_1^x = I_{in} \cdot \Big(\frac{2}{\pi}\Big)^2 \cdot sin^2(\pi f) \cdot e^{-\alpha d} \cdot sin^2(Fd)
\end{split}
\vspace{3mm}
\end{equation}
as there is a grating structure along the $x$-direction and no grating along the $y$-direction. Following this, the magneto-optical grating efficiency can be determined:

\begin{equation}
    \eta = \frac{I_1}{I_{in}} = \frac{4}{\pi^2} \cdot sin^2(\pi f) \cdot e^{-\alpha d} \cdot sin^2(Fd)
\end{equation}
This result is identical to the expression provided by Haskal\cite{Haskal_1970}. While there are various different approaches\cite{Fan_Hologram_1969,Haskal_1970,mezrich_reconstruction_1970}, the unit box function summation method is very versatile in defining the grid, which could be made to include finite-sized domain walls or mesoscopic magnetic textures, as well.

Finally, for small Faraday rotation angles the optimized thickness is $d = 2/\alpha$ and the maximum efficiency is thus given by:

\begin{equation}
    \eta_{max} = \frac{4}{\pi^2} e^{-2} \left( \frac{2F}{\alpha} \right)^2 sin^2(\pi f)
\end{equation}

\section*{Faraday spectra}
\hspace{\parindent}
In order to perform magneto-optical scattering measurements at the visible wavelength where the YIG magneto-optical activity is largest, we performed wavelength dependent measurements of the Faraday rotation\cite{Schmidt:2020de}. The measured spectrum is shown in Supplementary Fig.~\ref{Faraday_spectrum} highlighting the maximum response of the films in the green wavelength region. 

\begin{figure}[h!]
    \centering
    \includegraphics[width=0.5\columnwidth]{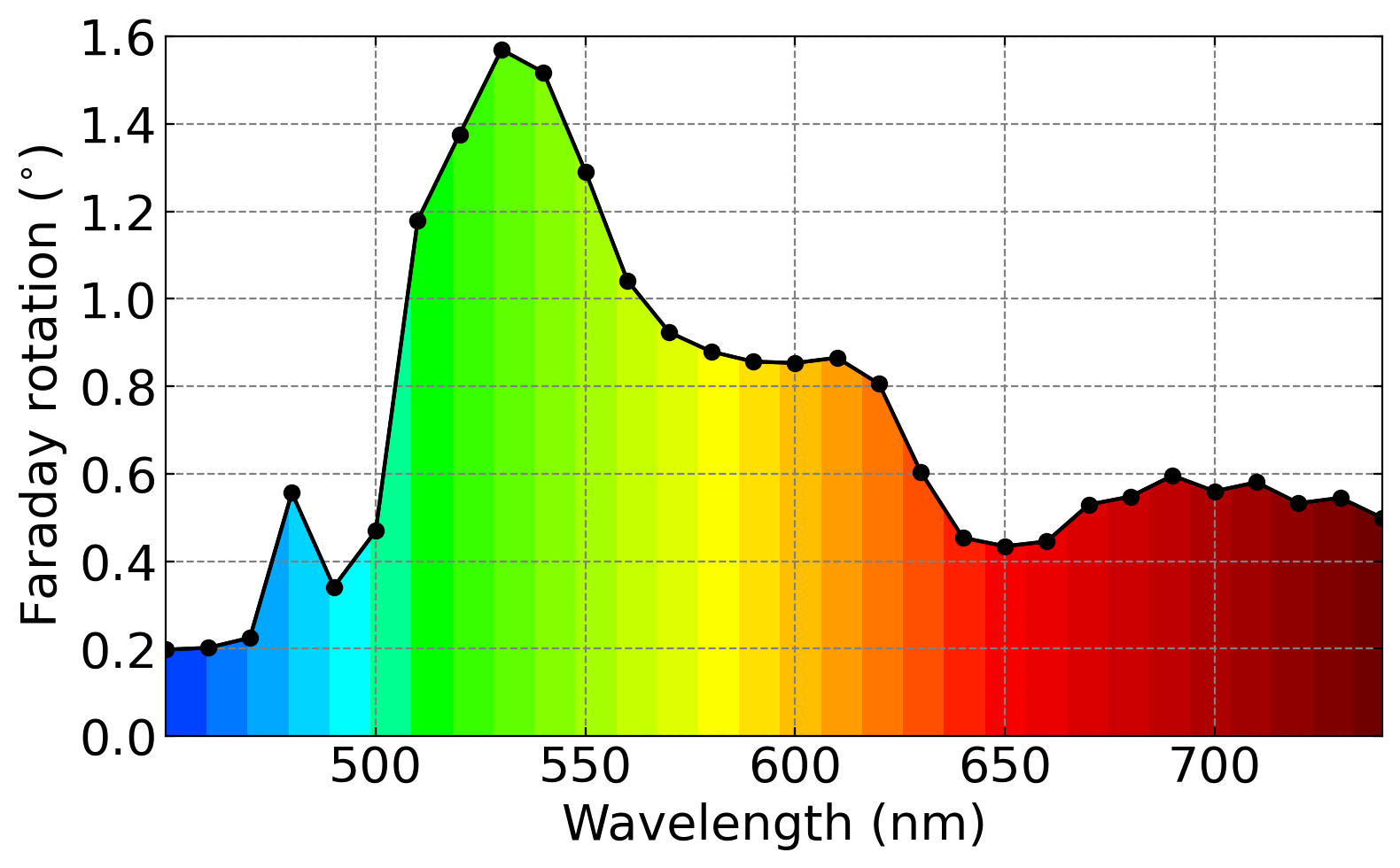}
    \caption{Faraday rotation of the YIG film. Spectra dependence of the Faraday rotation in the visible range. A distinct maximum for wavelengths corresponding to green wavelength can be seen. Based on this a laser wavelength of 530 nm was chosen for the scattering studies presented in the main text.}
    \label{Faraday_spectrum}
\end{figure}

A mercury lamp was used as a broadband white light source, which in conjunction with a Newport Cornerstone monochromator, allowed the wavelength to be swept in 10 nm increments. The output light was focused onto the YIG sample with polarization parallel to the short width of the YIG strip i.e. parallel to the stripe domains. The YIG sample was located within an electromagnet providing fields in the range -1.2  to 1.2 T. In the Faraday configuration the transmitted light is measured, and the Faraday rotation was extracted using an analyser - photo-elastic modulator (PEM) combination on the transmitted beam path, as outlined in the Hinds instruments PEM application note~\cite{Oakberg_2010}. A Hamamatsu H11901-20 photomultiplier tube was used as the photodetector. For each wavelength the out-of-plane magnetic field was swept between a saturating field of $\pm$ 300 mT and a hysteresis loop was recorded, from which the Faraday rotation was extracted as half the optical rotation between positive and negative magnetic saturated states.

\begin{figure}[t!]
    \centering
    \includegraphics[width=\columnwidth]{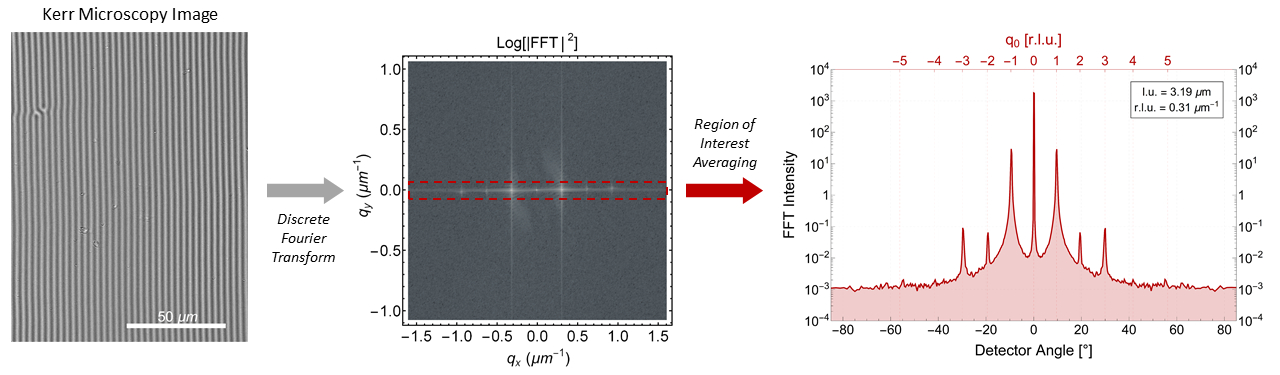}
    \caption{Real and reciprocal space. Graphical illustration of the connection between Kerr microscopy and computed reciprocal space patterns, presented in the main text. The example shown here relates a Kerr microscope image for the remanent magnetic state of the YIG film, with the resulting reciprocal space map resulting from a 2D Fourier transformation. The right panel illustrates the expected scattering pattern and related detector angles for a wavelength of $\lambda$~=~530~nm.}
    \label{Transform}
\end{figure}

\section*{Linking real- to reciprocal-space}

Utilizing the recorded magneto-optical Kerr effect images from our microscope, we computed the respective reciprocal space maps for all the applied magnetic field values. We further compared these to the actual recorded magneto-optical scattering patterns as shown in Fig. 4 of the main article text. Supplementary Fig. \ref{Transform} graphically depicts the process of computing the scattering patterns from the microscopy data.

\begin{figure}[t!]
    \centering
    \includegraphics[width=0.6\columnwidth]{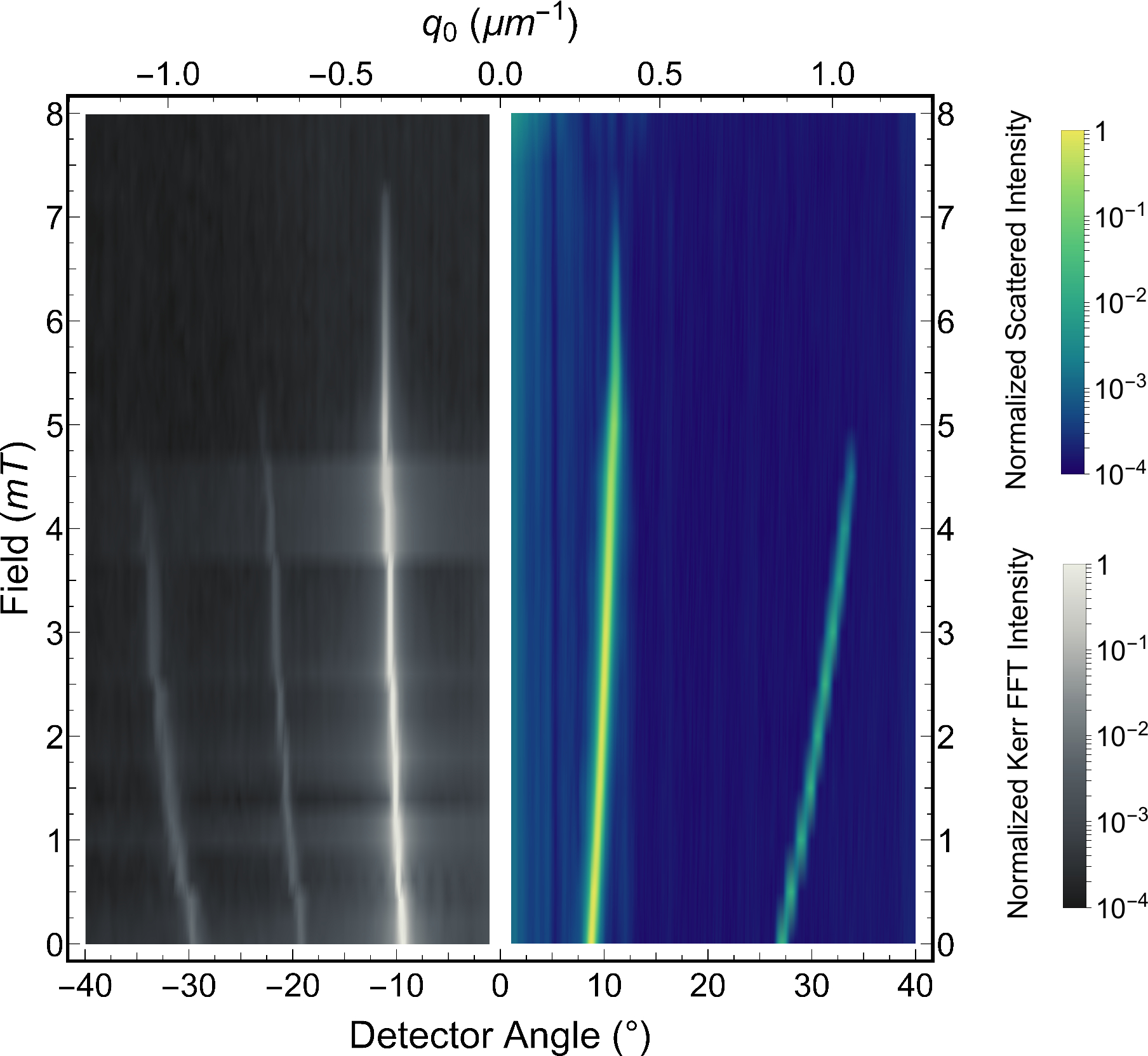}
    \caption{Field dependence maps of the magneto-optical scattering, determined by magnetic microscopy and magneto-optical scattering. Left: Maps computed from Kerr microscopy images for applied fields applied along the $y$ direction in the range of -8 to 8 mT. Right: Actual field dependence map resulting from magneto-optical scattering data.}
    \label{Maps}
\end{figure}

Starting from an in-plane saturated state, a series of Kerr microscopy images were recorded during a full hysteresis sweep, highlighting the evolution of the magnetic domain texture in an externally applied field. We performed a Discrete Fourier Transform for each of these images, obtaining reciprocal space maps for the corresponding magnetic domain textures, with distinctive peaks arising from the periodic features of the domain configuration and shown in Supplementary Fig. \ref{Maps}. Selecting an appropriate region of interest along the grid vector direction, a projected, one-dimensional reciprocal signal is generated, which can then be directly compared to the data obtained from the light scattering measurements. Furthermore, by tracking the position of the first order peaks as a function of applied field, see Supplementary Fig. \ref{Suppl Figure - Lattice Parameters}(a), the reciprocal lattice unit can be directly extracted. Note that there is an offset between the two field sweep directions, which is attributed to the remanence of the Kerr microscope's electromagnet poles. Assuming a symmetric linear response of the peak position with the applied field around remanence, a fact also confirmed by the light scattering measurements (see inset of Fig. 4 of the main text), an absolute value fit function is used to extract the field offsets. The fitting window is centered around the directly recorded remanence and its extent was defined by taking the field span for which the best overall match was achieved for the three fitting parameters using the two datasets independently.

The real lattice parameters are obtained by inverting the reciprocal lattice units and their dependence on the applied fields is represented Supplementary Fig. \ref{Suppl Figure - Lattice Parameters}(b). The step-like shape of both plots of Supplementary Fig. \ref{Suppl Figure - Lattice Parameters} is reminiscent of the discrete nature of the peak positions, measured in number of pixels.

\begin{figure}[t!]
    \centering
    \includegraphics[width=\columnwidth]{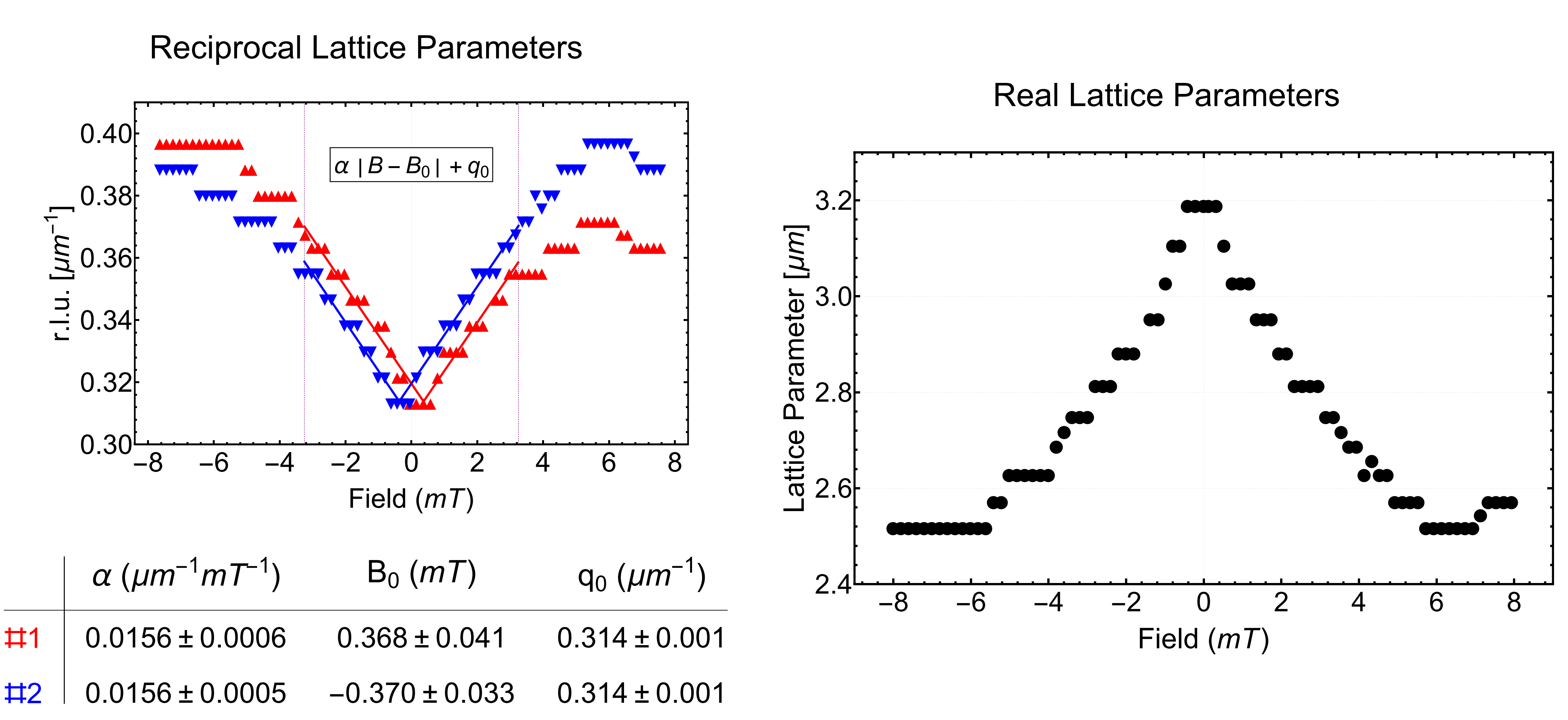}
    \caption{(a) The reciprocal lattice unit (r.l.u) dependence on the applied field for the two field sweeps, going from negative to positive saturation (red, up-triangle data) and from positive back to negative saturation (blue, down-triangle data). The inset provides the fitting function used, while the table contains the fit parameters extracted for each sweep, respectively. The vertical purple lines mark the extent of the fitting window. (b) The evolution of the real lattice parameter as a function of applied field.}
    \label{Suppl Figure - Lattice Parameters}
\end{figure}

\begin{figure}[h!]
    \centering
    \includegraphics[width=0.5\columnwidth]{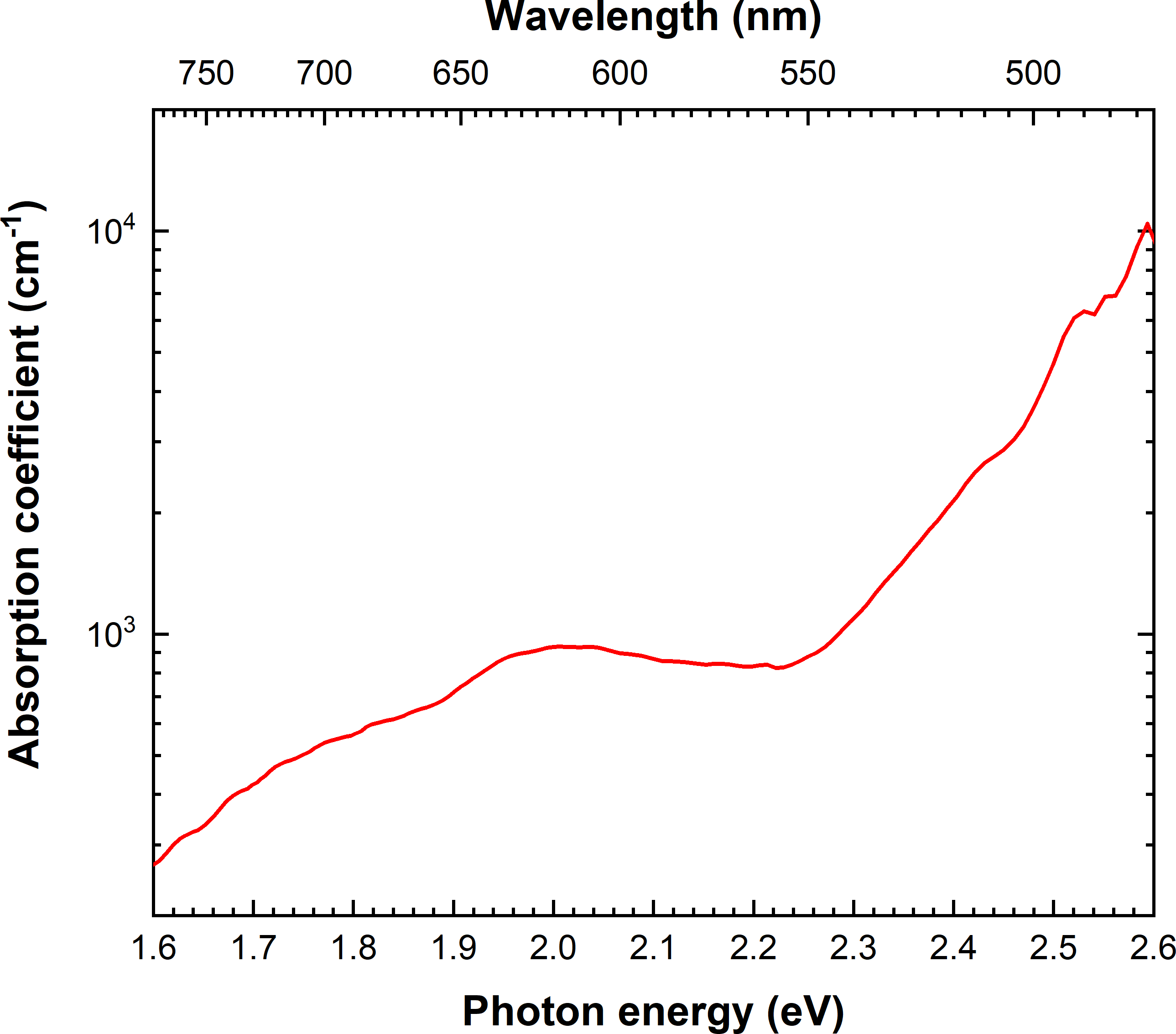}
    \caption{Optical absorption coefficient of the YIG film studied in this work.}
    \label{fig:absorption}
\end{figure}

\section*{YIG optical absorption}

Supplementary Fig. \ref{fig:absorption} shows the optical absorption coefficient for the YIG sample. Values of the coefficient from this dataset were used for the theoretical estimates of maximum grating efficiency discussed in the main text.

\newpage
\section*{References}

\end{document}